\documentclass[referee]{raa}           

\usepackage{graphicx,times}
\usepackage{diagbox}
\usepackage{bm}
\usepackage{natbib}
\usepackage{makecell}
\usepackage{fontenc}
\usepackage{float}
\usepackage{subcaption}
\usepackage{amssymb,amsmath}
\bibpunct{(}{)}{;}{a}{}{,}

\usepackage{makecell}
\usepackage{array}
\usepackage[pagebackref=true]{hyperref}

\begin{document}

   \title{Applying hybrid clustering in pulsar candidate sifting 
   	with multi-modality for FAST survey
   }

 \volnopage{ {\bf 20XX} Vol.\ {\bf X} No. {\bf XX}, 000--000}
   \setcounter{page}{1}

   \author{Zi-Yi You  
   \inst{2,5}, Yun-Rong Pan\inst{2}, Zhi Ma
   \inst{2}, Li Zhang()\inst{1,3}, Shuo Xiao\inst{2,5}, Dan-Dan Zhang\inst{2}, Shi-Jun Dang\inst{2,5}, Ru-Shuang Zhao\inst{2,4}, Pei Wang\inst{4}, Ai-Jun Dong\inst{2,5} , Jia-Tao Jiang\inst{8}, Ji-Bing Leng\inst{6}, 
      Wei-An Li\inst{6}, Si-Yao Li\inst{7}
      }

   \institute{Corresponding author lizhang. {\it science@gmail.com}\\
        \and
              School of Physics and Electronic Science, Guizhou Normal University, Guiyang 550025, China\\
	\and
College of Big Data and Information Engineering, Guizhou  University, Guiyang 550025, China \\
\and 
CAS Key Laboratory of FAST, National Astronomical Observatories, Chinese Academy of Sciences, Beijing,100101, China\\
\and
Guizhou Provincial Key Laboratory of Radio Astronomy and Data Processing, Guizhou Normal University, Guiyang 550025, China\\
\and
School of Big data and Computer Science, Guizhou Normal University, Guiyang 550025, China\\
\and
Guizhou Software Engineering Research Center,Guiyang 550000, China\\
\and
Key Laboratory of Information and Computing Guizhou Province, Guizhou Normal University, Guiyang 550001, China\\
\vs \no
   {\small Received 2023   September 8; accepted 2023  November 10}
}

\abstract{ Pulsar search is always the basis of pulsar navigation, gravitational wave detection and other research topics. Currently, the volume of pulsar candidates collected by Five-hundred-meter Aperture Spherical radio Telescope (FAST) shows an explosive growth rate that has brought challenges for its pulsar candidate filtering System. Particularly, the multi-view heterogeneous data and class imbalance between true pulsars and non-pulsar candidates have negative effects on traditional single-modal supervised classification methods. In this study, a multi-modal and semi-supervised learning based pulsar candidate sifting algorithm is presented, which adopts a hybrid ensemble clustering scheme of density-based and partition-based methods combined with a feature-level fusion strategy for input data and a data partition strategy for parallelization. Experiments on both HTRU (The High Time Resolution Universe Survey) 2 and FAST actual observation data demonstrate that the proposed algorithm could excellently identify the pulsars: On HTRU2, the precision and recall rates of its parallel mode reach 0.981 and 0.988. On FAST data, those of its parallel mode reach 0.891 and 0.961, meanwhile, the running time also significantly decrease with the increment of parallel nodes within limits. So, we can get the conclusion that our algorithm could be a feasible idea for large scale pulsar candidate sifting of FAST drift scan observation.
}

   \authorrunning{Z.-Y. You et al. }            
   \titlerunning{Applying hybrid clustering in pulsar candidate sifting }  
   \maketitle

%
\section{Introduction}           

So far, a lot of radio pulsars have been discovered by the modern pulsar surveys, including High Time Resolution Universe (HTRU, \citealt{Burke-Spolaor+etal+2011}) Parkes survey, the low-frequency array (LOFAR) tied-array all-sky survey (LOTASS, \citealt{Coenen+etal+2014}),
the Commensal Radio Astronomy FasT Survey (CRAFTS, \citealt{Jiang+etal+2019}, \citealt{wang+etal+2021}) and Galactic Plane Pulsar Snapshot (GPPS, \citealt{han+2021+fast}) etc. Compared to previous ones, modern surveys tend to  use more sensitive detection techniques, greater survey areas, improved data analysis techniques and collaborative efforts, such as CRAFTS
. They often produce a large number of potential pulsar candidates, but only a very small proportion of these candidates (nearly one in ten thousand) are identified as real pulsars for the reason of large amounts of interference signals. Thus, it is critical to reduce the retention of a large number of  non-pulsar signals without loss of pulsar-like samples. At present, this issue can be alleviated on two research points of pulsar search pipelines. i) signal processing: remove Radio Frequency Interference (RFI) signals from a large amount of observational data as much as possible (\citealt{Morello+etal+2014}, \citealt{yang+etal+2020}) and optimize some important parameters such as signal-to-noise ratio (S/N) detections and so on. ii) candidate selection: minimize the labor of further observations, which focuses on filtering the pulsar-like samples among large numbers of candidates automatically and accurately by advanced artificial intelligence techniques. This paper involves the latter.

Generally, existing pulsar candidate selection methods based on artificial intelligence could be classified into three categories, according to the principles of these methods. The traditional scoring methods are regarded as the first category e.g.,(PEACE, \citealt{Lee+etal+2013}). The second category refers to the Machine Learning (ML) based classifiers that usually perform better than the first category, e.g., (\citealt{Morello+etal+2014}), (\citealt{Lyon+etal+2016}) and (\citealt{Tanc+etal+2018}), etc. More recently, (\citealt{Chakraborty+2019}) applied the eight features designed by (\citealt{Lyon+etal+2016}) to test the performance of Random Forest (RF) algorithm, K-Nearest Neighbors(KNN) algorithm and Logistic Regression (LR) algorithm. (\citealt{xiao+etal+2020}) designed a reliable KNN based model named Pseudo-Nearest Centroid Neighbour classifier (PNCN) for pulsar survey data streams, which can effectively deal with the class imbalance problem. In these methods, the features used often depend heavily on human experience, whose classification performance may be adversely affected. For example, some classifiers only extract features from the Dispersion Measurement (DM) and pulse profile curve, which leads to some RFIs being identified as pulsars incorrectly. As the radio environment is becoming more complex, it is more difficult to effectively distinguish pulsar candidates and non-pulsar candidates only by statistical features. In practice, the pulsars can be successfully identified just by human experts observing the corresponding diagnostic plots. Based on this inspiration, the third category utilizes diagnostic plots as the inputs of image recognition models or multi-method ensemble models including image recognition, so that the "pulsar-like" mode can be learned from the diagnostic sub-graph automatically by training the Deep Learning (DL) based models, e.g., (\citealt{wang2019pulsar}), (\citealt{Guo+etal+2019}), (\citealt{zeng+2020}) and (\citealt{zhang+etal+2021}) etc. Compared with the ML based models of second category, these methods have better generalization ability. Among them, the methods (\citealt{wang2019pulsar}) and (\citealt{zeng+2020}) are mainly used in the pulsar search pipeline of FAST survey. (\citealt{wang2019pulsar}) presented a new ensemble classification system on FAST for pulsar candidate selection that composed of five classifiers. Furthermore,it denotes the development of Pulsar Image-based Classification System (PICS) (\citealt{zhu+etal+2021}). (\citealt{zeng+2020}) designed an end-to-end online learning model, namely Concat Convolutional Neural Network (CCNN) to identify the candidates without any intermediate labels which are processed from FAST data. 
The aforementioned methods are mostly based on single mode. At present, there are few literatures about multi-modal methods applied to astronomical data mining especially pulsar identification. (\citealt{zhang+etal+2021}) proposed an   early fusion based pulsar image identification framework with smart under-sampling, which was evaluated on HTRU medlat dataset. In this work, a semi-supervised learning and Feature-level Multi-modal Fusion based Hybrid Clustering scheme (FMFHC) is designed for large scale candidate sifting through FAST pulsar search pipelines, in terms of (\citealt{wang2019pulsar}) and (\citealt{MA+etal+2022}).

A typical pulsar search pipeline for FAST drift scan observed data roughly includes following several steps: i) eliminating the obvious interference signals from original data ; ii) dedispersing the data into time series with distinct DM values; iii) performing a fast Fourier transform on each time series so as to further search for periodic signals; iv) sifting these periodic signals and outputting candidates obtained to files (e.g. suffix pfd); v) folding the data in a periodic manner, and then output candidate images. In practice, there are still a lot of non-pulsar candidates in step iv), including RFI. Our algorithm is introduced into the sifting stage of step ,iv) aiming to further address the following issues in the pulsar candidate selection area:

\textbf{i)For some aforementioned methods (e.g. the supervised ML based candidate signal classifiers and DL based diagnostic subplots recognition models), the cost of obtaining a large amount of labeled data (in which the proportion between the real pulsar and the non-pulsar sample is extremely imbalanced) and periodic training (to avoid overfitting and underfitting) is too high, and these methods are all based on binary classification }.

\textbf{ii)In the process of pulsar search, there are usually multi-view heterogeneous candidate data, which contain various types and attributes. In practical applications, it could be difficult to further mine the deep features hidden in these data through a single modal candidate selection algorithm}.

The rest of this paper is organized as follows: Section 2 describes the pulsar candidate features and similarity measure involved. Section 3 presents the components of overall algorithm in detail. In Section 4, the experimental data sets, data pre-processing methods and results are illustrated. The discussion are in Section 5. Finally, in Section 6, we present the conclusion and the future work.

\section{Pulsar Candidate Features and Similarity Measure}
\label{sect:Mea}
\subsection{Pulsar Candidate Features}


The extraction of candidate features is very important to maximize the separation between non-pulsar and pulsar candidates. Suppose the pulsar candidates we considered were processed by the software pipelines based on PulsaR Exploration and Search Toolkit (PRESTO, \citealt{ransom+2011}, \citealt{Yue+2013}), which implemented the similar search steps for advanced telescope systems such as FAST.
\begin{table}[H]
	\bc
	\begin{minipage}[]{100mm}
		\caption[]{The detailed information on the plots in Figure 1}\end{minipage}
	\setlength{\tabcolsep}{2.5pt}
	\small
	\begin{tabular}{c|cccccccccccc}
		\hline\noalign{\smallskip}
		\diagbox{parameters}{Name}  &  J0358+5413 &  J1915+1606&interference signal I&interference signal II \\
		\hline\noalign{\smallskip}
		Telescope:&   FAST &  FAST& FAST&FAST  \\
		Candidate:& 156.38 ms\_cand& 59.03 ms\_cand& JERK $\_$ cand 30 & JERK $\_$ cand 35 \\
		Epoch$_{topo}$:&59435.95677812153&58784.35905500046&58419.36111111111&58419.36111111111 \\
		Epoch$_{bary}$:&59435.95439208478&58784.35452882630&58419.36110676074&58419.36110598373 \\
		Data folded:&2620416&1309696&12057600&12057600 \\
		Data Avg: &2.387e+04&1.103e+05&1.48e+05&1.48e+05 \\
		Data StdDev: &332.9&569.7&782.1&782 \\
		Profile Bins:&64&64&64&64 \\
		Profile Avg:&9.772e+08&2.257e+09&2.787e+10&2.787e+10 \\
		Profile StdDev:&6.737e+04& 8.15e+04&3.395e+05&3.394e+05 \\
		DOFeff:&59.53&60.31&59.00&59.53 \\
		DM(pc/cm$^{3}$)&57.727&168.770&203.538&242.389 \\
		Ptopo(ms):&156.3697468&59.027867&17.3762014&22.3537411 \\
		P$^{'}$topo(s/s):&$4.602(20) \times 10^{-8}$&$0.0(1.0)\times 10^{-8}$&$ 2.3313(78)\times 10^{-8}$&$9.78(10)\times 10^{-9}$ \\
		P$^{''}$topo(s/s$^{2}$):&$ 0.0(5.0)\times10^{-12}$&$ 0.0(1.0)\times10^{-9}$&$ 0.0(8.6)\times10^{-13}$&$ 0.0(1.1)\times10^{-12}$ \\
		P$_{Noise}$:&$~0(6207.2\sigma)$&$~0(313.5\sigma)$&$<4.18e-09(5.8\sigma)$&$<2.31e-13(7.2\sigma)$ \\
		P$_{bary}$(ms):&156.3274369(66)&59.030003(89)&17.3762014(60)&22.3537411(79) \\
		Ptopo(ms):&156.3697468&59.027867&17.3762014&22.3537411 \\
		P$^{'}$topo(s/s):&$4.604(20) \times 10^{-8}$&$0.0(1.0)\times 10^{-8}$&$ 2.3313(78)\times 10^{-8}$&$9.78(10)\times 10^{-9}$ \\
		P$^{''}$topo(s/s$^{2}$):&$ 0.0(5.0)\times10^{-12}$&$ 0.0(1.0)\times10^{-9}$&$ 0.0(8.6)\times10^{-13}$&$ 0.0(1.1)\times10^{-12}$ \\
		RA$_{J2000}$:&12:34:56.7890&12:34:56.7890&12:34:56.7890&12:34:56.7890 \\ 
		DEC$_{J2000}$& -12:34:56.7890&-12:34:56.7890&-12:34:56.7890 &-12:34:56.7890 \\ 
		Receiver Name:&19-beam receiver&19-beam receiver&19-beam receiver&19-beam receiver \\ 
		Antenna Gain:& 16.1k Jy-1& 16.1k Jy-1& 16.1k Jy-1& 16.1k Jy-1 \\ 
		Bandwidth:&500MHz&500MHz&500MHz&500MHz \\ 
		Temperature:&~20K&~20K&~20K&~20K \\ 
		Terminal Name: &psr&psr&psr&psr \\ 
		Sampling Rate:&49.152 $\mu$s&49.152 $\mu$s&49.152 $\mu$s&49.152 $\mu$s \\
		Polarization channels:&4&4&4&4 \\
		Number of channels:&4096&4096&4096&4096 \\
		
		\noalign{\smallskip}\hline
	\end{tabular}
	\ec
\end{table}
\begin{figure}[htbp]
	\centering
	\subfloat[]
	{\includegraphics[width=0.5\textwidth]{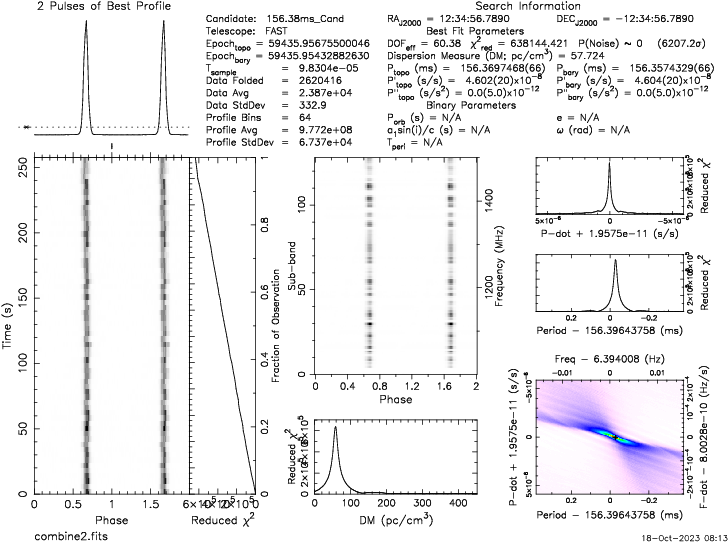}\label{fig:subfig7}}
	\subfloat[]
	{\includegraphics[width=0.5\textwidth]{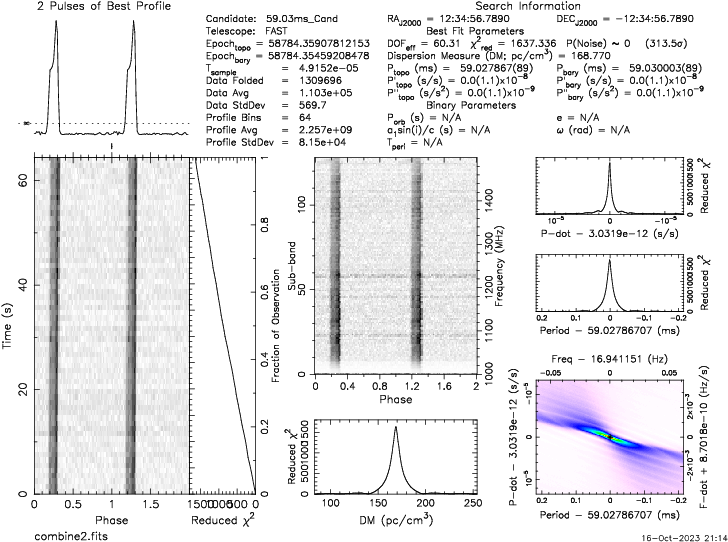}\label{fig:subfig8}}
	
	\subfloat[]
	{\includegraphics[width=0.5\textwidth]{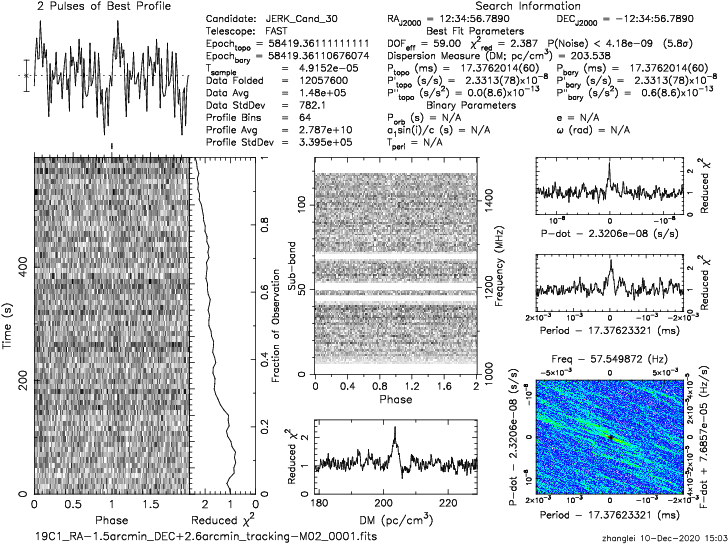}\label{fig:subfig9}}
	\subfloat[]
	{\includegraphics[width=0.5\textwidth]{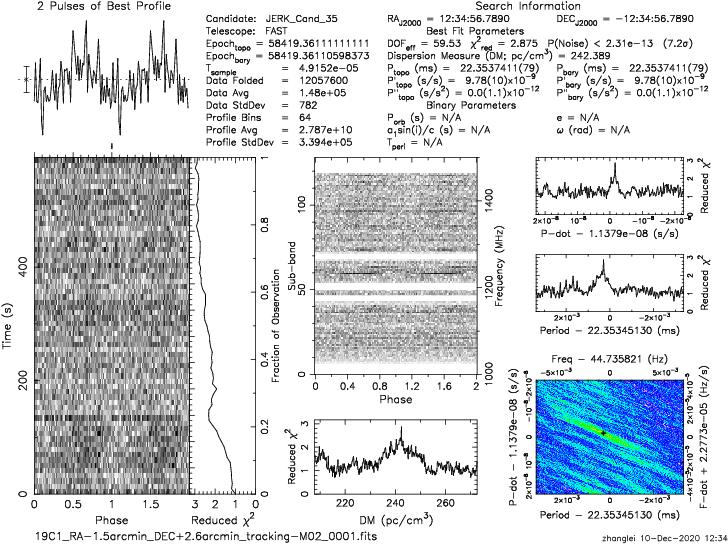}\label{fig:subfig10}}
	
	\caption{ An example of the similarity between candidates. Note that, (a) is the diagnostic plots of known pulsar J1906+0746 and (b) is those of known pulsar J1807-0847; (c) is the diagnostic plots of interference signal I and (d) is those of interference signal II.}
\end{figure}

In terms of statistical features, there are $8$ new features extracted from the pfd files by feature extraction program (\citealt{Lyon+etal+2016}), including the mean value, excess kurtosis, standard deviation and skewness of pulse profile, and the mean value, excess kurtosis, standard deviation and skewness of the DM-SNR (Signal-to-Noise Ratio) curve. Note that, the first four statistics correspond to the integrated pulse profile, and the remain four correspond to the DM-SNR curve. These features were chosen to maximize the separation of various candidate classes when used together with ML classifier. Furthermore, the HTRU2 data set was used during the work (\citealt{Lyon+etal+2016}). In terms of diagnostic plots, most features of a candidate signal can be visualized through different diagnostic subplots, including folded profile plot, sub-integrations plot, sub-bands plot, and DM-SNR curve, etc. The ensemble model based on PICS (\citealt{wang2019pulsar}) can also extract four main feature plots of a candidate from the pfd files, which are one-dimensional (1-D) data arrays summed profile and DM curve, two-dimensional (2-D) data arrays time versus phase (TVP) and frequency versus phase (FVP). Note that, the size of 1-D feature plots is 64×1, while the size of 2-D feature plots is 64×64. The experiments implemented on FAST pulsar survey data (\citealt{wang2019pulsar}) demonstrate that, PICS-ResNet can achieve a higher recall rate of 98 percent than PICS (that is 95 percent).

\subsection{Similarity Measure}

A candidate believed to be a real pulsar must have very similar statistical features (e.g.standard deviation and skewness of pulse profile) and diagnostic subplots (e.g.2-D FVP) with some other known pulsars. That is the reason we plan to design a multi-modal clustering algorithm for large mounts of pulsar candidate data. Fig.1 shows an example of the features and plots. It can be  seen that known pulsars J0358+5413 and J1915+1606 are very similar in the integrated pulse profile and frequency versus phase diagram. Similarly, interference signal I and interference signal II are also very similar in the integrated pulse profile and frequency versus phase diagram. The detailed information on these plots are described in Table 1. The feature similarity between candidates will be further validated in experiment results of Section 4.2, as shown in Fig.8.

\section{The Method}
\label{sect:Obs}
\subsection{Feature Fusion Strategy}
\label{sect:Obs}
\begin{figure}[H]
	\centering
	\includegraphics[width=12cm]{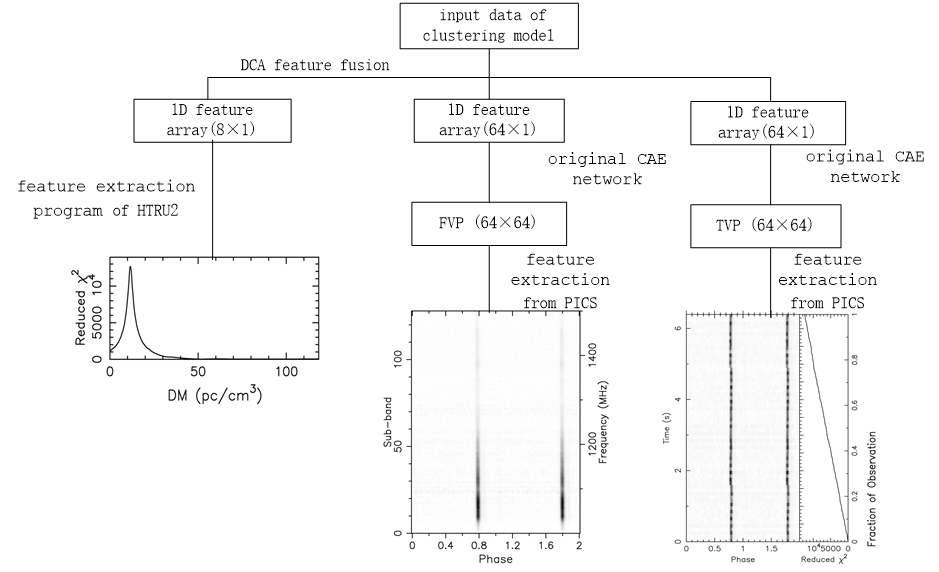}
	\caption{ Diagram of the feature fusion model.  } 
	\label{Fig2}
\end{figure}

Through fusing different features of multiple modalities extracted from a single candidate, feature fusion methods can further refine the features with higher discrimination. Among them, the Discriminant Correlation Analysis  (DCA) is a linear method which maximizes the pair-wise correlation between two feature sets and possesses very low computational complexity at the same time.

Depending on DCA algorithm, our objective is pre-processing candidate data extracted from pfd files by fusing features from different modalities of the same object before feeding them to hybrid clustering scheme as input data. According to Section 2.1, these modalities include 1-D data array (statistical feature format on HTRU2) and 2-D arrays (FVP and TVP formats on PICS), as shown in Fig.2. DCA algorithm is able to establish the correlation criterion between the two groups of feature vectors, to extract their canonical correlation features. In this work, it is assumed that $N$ training samples are collected from 2 classes, which are \{0: non-pulsar, 1: pulsar\}. For each sample, two feature vectors with 8 and 64 dimensions are extracted from two modalities, which are 1-D statistical features extracted from the feature extraction program of HTRU2 and 2-D feature plots extracted from PICS. Then, $X = R^{8\times N}$ and $Y = R^{64\times N}$ denote the data matrices containing the two feature sets. $X$ template is composed of $N$ 1-D vectors (8×1) which have the same format as that of HTRU2. Furthermore, $Y$ template is generated by extracting $N$ feature plots (TVP or FVP or fusion of both, $64 \times 64$) using the unsupervised Convolutional Auto-Encoder network (CAE) as shown in Fig.3. Note that, each feature plot is dimensionally reduced to $8 \times 8$ through CAE and then resized to $64 \times 1$ by the $reshape$ method in Python toolkit. So, the dimension of $Y$ is defined as $64 \times N$.
 
\begin{figure}[H]
	\centering
	\includegraphics[width=12cm]{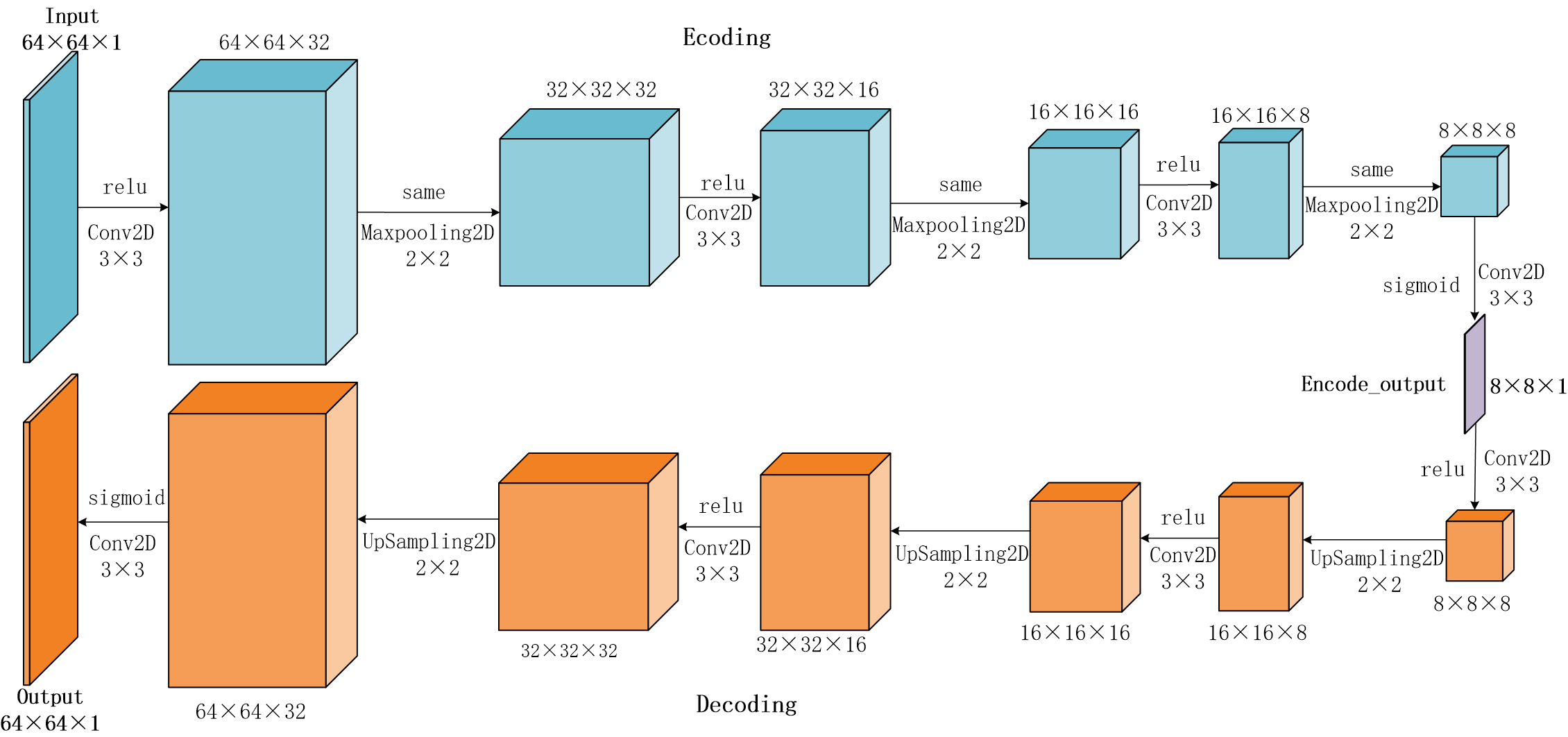}
	\caption{ Network architecture of original CAE. Note that, the encoding module of CAE has 4 hidden layers as follows. The first layer :1 2-D convolution with $(3 \times 3) \times 32$ (output channels) and 1 2-D max-pooling with $2 \times 2$; the second layer: 1 2-D convolution with $(3 \times 3) \times 16$ and 1 2-D max-pooling with $2 \times 2$; the third layer: 1 2-D convolution with $(3\times 3)\times 8$ and 1 2-D max-pooling with $2 \times 2$ ; the fourth layer: 1 2-D convolution with $(3 \times 3) \times 1$. The following value was used:the Input size, $64 \times 64 \times 1$; kernel, $3 \times 3$ pixels; padding, $1 \times 1$ pixels; the Encode\_output size, $8 \times 8\times 1$.} 
	\label{Fig3}
\end{figure}

The fine tuned CAE architecture we used is shown in Fig.3. Then, after model training (where Epochs = 50, Batch\_size
 =128, the optimizer is Adadelta and the loss function is Binary\_crossentropy) and testing, the loss rate was maintained at around $35\%$, which can make the model retain most of the main features of TVP or FVP while ensuring runtime efficiency. By using this CAE, the entire clustering algorithm performed well in the subsequent experiments in Section 4.1 and 4.2. Meanwhile, the process of DCA method is described as follows:
 
i)The $N$ columns of both data matrix are divided into $c$ separate groups. Let $S_{bx}$   be the between-class scatter matrix defined as Eq(1) and Eq(2):
\begin{equation}\label{eq1}
		S_{bx_{(8\times 8)}} = \sum_{i=1}^{c} n_{i}(\overset{-}x_{i}-\overset{-}{x}
	)(\overset{-}x_{i}-\overset{-}x) ^{T} = \Phi
	_{bx_{(8\times c)}}\Phi_{bx_{(c\times 8)}}^{T}
\end{equation}
\begin{equation}\label{eq2}
	\Phi_{bx_{(8\times C)}} = [(\overset{-}x_{1}-\overset{-}{x}),\sqrt{n_{2}}
	(\overset{-}x_{i}-\overset{-}x),...,\sqrt{n_{c}}(\overset{-}{x_{c}}-\overset{-}{x})]
\end{equation} 
where $x_{ij} \in X$ denotes the p dimensional feature vector of the $j^{th}$ sample in the $i^{th}$ class, $\overset{-}{x_{i}}$ denotes the mean of the $i^{th}$ class, and $\overset{-}{x}$ denotes the whole feature set mean.

ii)Finding transformation matrix $W_{bx}$  to transform the corresponding feature matrix $X_{8\times N}$  to $X_{r\times N}^{'}$  as follows:
\begin{equation}\label{eq3}
X_{(r\times n)}^{'}=W_{bx(r\times 8)}^{T}X_{(8\times n)}
\end{equation}
\begin{equation}\label{eq4}
	S_{bx}^{'} = W_{bx}^{T}S_{bx}W_{bx} = \Phi_{bx}^{'}\left(\Phi_{bx}^{'}\right)^{T} = I
\end{equation}
where $\Phi_{bx}^{'}\left(\Phi_{bx}^{'}\right)^{T}$ is a strictly diagonally dominant matrix, in which  denotes the correlation between the $i^{th}$ class and the $j^{th}$ class. Similar to $X_{8\times N}$ , finding transformation matrix $W_{by}$	 to unitize the between-class scatter matrix $S_{by}$ , which transforms $Y_{64\times N}$ to  $Y_{r\times N}^{'}$ .

iii)Maximizing the pair-wise correlation across the feature sets $X^{'}$ and $Y^{'}$. It requires the between-set covariance matrix $(S_{xy}^{'}= X^{'}Y^{'T})$  to be diagonal through singular value decomposition (SVD) as follows:
\begin{equation}\label{eq5}
	S_{xy_{(r\times r)}}^{'} = U\Sigma V^{T} \Rightarrow (U\Sigma^{-\frac{1}{2}
	})^{T}S_{xy}^{'}(V\Sigma^{-\frac{1}{2}})T = I
\end{equation}

The transformation matrices for $X^{'}$ and $Y^{'}$  are:
\begin{equation}\label{eq6}
W_{cx} = U\Sigma^{-\frac{1}{2}}
\end{equation}
\begin{equation}\label{eq7}
	W_{cy} = V\Sigma^{-\frac{1}{2}}
\end{equation}

Consequently, the final transformed feature sets are:
\begin{equation}\label{eq8}
X^{*} = W^{T}_{cx}W_{bx}^{T}X
\end{equation}
\begin{equation}\label{eq9}
	Y^{*} = W^{T}_{cy}W_{by}^{T}Y
\end{equation}

iv)The early feature-level fusion is performed by the summing of the final transformed feature sets, that is $Z = X^{*}+Y^{*}$, where $Z$ called the Canonical Correlation Discriminant Features (CCDFs).

In this way, the data samples with fused features can be inputted into the hybrid clustering model in the next step.

\subsection{Hybrid Clustering}
\begin{figure}[H]
	\centering
	\includegraphics[width=9cm]{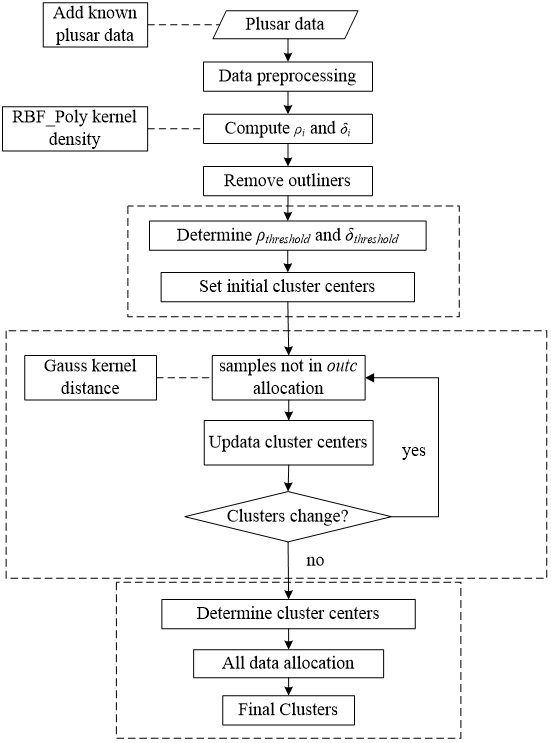}
	\caption{Flow chart of hybrid clustering scheme.} 
	\label{Fig4}
\end{figure}

Aiming to classify data into various homogeneous clusters in terms of their similarities, clustering methods are divided into different categories including density-based methods, partition-based methods and so on. As a representative partition-based clustering, KMeans is widely applied (\citealt{Krishna+1999}). Due to its drawbacks (it is only applicable to distinguish clusters with hyper spherical data distribution and the clustering results are usually sensitive to the parameter settings), the extension versions have been presented such as (\citealt{arthur+2007}) and (\citealt{Nguyen+2018}). In addition, the Density Peaks Clustering (DPC) algorithm adopted density peaks as features to quickly discover the potential cluster centers without any prior knowledge through drawing a two-dimensional decision graph. More recently, another density-based algorithm using global and local consistency adjustable manifold distance (McDPC) was proposed by (\citealt{wang+2020+mcdpc}) to address the drawback (that it is easy to divide the same cluster into multiple micro-clusters corresponding to its multiple high-density points) of CFSFDP (\citealt{rodriguez+2014}). FMFHC is developed as a fusion of clustering methods based on DPC and KMeans.

The clustering process of FMFHC is summarized in Fig.4, in which the main steps are concluded as follows:

i)To better adapt to different data structures, the k-nearest neighbor based mixed kernel function of RBF and Polynomial (RBF\_Poly) is used to calculate the density values of fused input data (mentioned in Section 3.1), as shown in Eq.(10) and Eq.(11).
\begin{equation}\label{eq10}
K_{RBF\_Poly}(x_{i},y_{j}) = \lambda \ exp[-\frac{||x_{i}-x_{j}||^{2}}{2\sigma^{2}}]+(1-\lambda)(x_{i}\cdot y_{j})((x_{i}\cdot y_{j})+1)^{q-1}, q \geq 1
,\lambda \in [0,1]
\end{equation}
where $K_{RBF\_Poly}(x_{i},y_{j})$ represents the mixed kernel function of  data points $x_{i}$ and $x_{j}$, $\lambda$  represents the weight of RBF function, $\sigma$ is the width of RBF function and $q$  is the order of the polynomial function. Although $\sigma$, $q$ and $\lambda$ cannot significantly improve clustering effect, they can make mixed kernel distance based similarity calculation more stable for multiple shapes of data distribution if $\lambda$ value is reasonable. The values of these three parameters were determined ( $\sigma=1, q =2,\lambda=0.95$) by past experiences since there is no analytical method for the selection of fusion coefficients currently, mentioned in (\citealt{wang+etal+2017}) and (\citealt{Huang+etal+2013}).

\begin{equation}\label{eq11}
\rho_{i} = \sum_{x_{j} \in KNN_{(x_{i})}} K_{RBF\_Poly}(x_{i},x_{j})
\end{equation}
where $\rho_{i}$  denotes local density of data  point $x_{i}$  and $KNN(x_{i})$  denotes the k-nearest neighbors of $x_{i}$ . Further, parameter $\delta_{i}$   of data point $x_{i}$  is defined as Eq.(12):
\begin{equation}\label{eq12}
	\delta_{i} = \min_{x_{j}:\rho_{j} \succ \rho_{i}}  S_{Mah}(x_{i},x_{j})
\end{equation}
where $S_{Mah}(x_{i},x_{j})$ denotes the Mahalanobis distance between $x_{i}$  and $x_{j}$. In addition, the density threshold $\rho_{outlier}$ is used for extracting outliers from the whole data set, some  of which may be special pulsars need to be further determined. An improved K-dist graph method is employed to determine $\rho_{outlier}$  (\citealt{wang+2020+mcdpc}). In the K-dist graph, outliers often are located at the leftmost local density level named \textbf{$PL$},  which comprises edge points of natural clusters with lower $\rho$ and lower $\delta$ values and outliers with lower  $\rho$ and higher $\delta$  values. To further distinguish which data points in \textbf{$PL$} should be considered as outliers, $\rho_{outlier}$  is determined as Eq.(13)
\begin{equation}\label{eq13}
	PL \supseteq \begin{cases}
		
			 \left\{ \xi:PL_{1}, \ldots, PL_{j} \right\}, \rho_{x_{1}}, \ldots, \rho_{x_{j}} > \rho_{outlier}\\
	\left\{ \text{$outc:$} PL_{j+1}, \ldots, PL_{w} \right\}, \rho_{j+1}, \ldots, \rho_{n} \leq \rho_{outlier}
	\end{cases}
\end{equation}
where $ PL_{j}$ denotes a data point in \textbf{$PL$}, \textbf{$outc$} denotes the set of outliers and \textbf{$outc$} are set in an autonomous manner, i.e. $\forall x_{i} \in outc$ , $\forall x_{j} \in \xi$ , $5\rho_{x_i} \leq \rho_{x_j}$.

ii)The improved cluster center selection scheme of DPC is used with automatic determining the number of clusters and their center points. All the values of $\rho_{i}$ and $\delta_{i}$ $(x_{i} \notin outc)$ are used for generating the two-dimensional decision graph which helps to select the initial cluster centers automatically. In the derived decision graph as shown in Fig.5, the parameters $\rho_{threhold}$ and $\delta_{threhold}$ are reasonably set as the truncation threshold to form a rectangle with a red border, in which all data points are selected as representative points (i.e. initial cluster centers). Moreover, the gap area with the yellow background separates these representative points from other points. Note that, the representative points are also the multi-density center points, the rest of data points will be allocated to these center points to form intermediate micro-clusters. This representative point selection scheme is similar to that in McDPC algorithm, which can divide the remaining samples into different density levels. But it has better generalization performance and can identify more multi-density datasets compared to McDPC.

iii)After the number of clusters $k$ and the initial cluster centers were determined, the improved iterative optimization scheme of cluster centers of KMeans is used for all data points regrouping and final convergence. The distance between any point $x_{i}$ $(x_{i} \notin outc)$  and each cluster center of current iterative is calculated based on RBF kernel function. Note that, RBF can improve the similarity measure between two points by mapping from measure distances to a high-dimensional space. Moreover, starting from the $2^{th}$ iteration, a weighted distance optimization is adopted for similarity measure as shown is Eq.(14) and Eq.(15):
\begin{equation}\label{eq14}
new\_\rho_{centerj} =\frac{Max\_\rho - Min\_\rho}{Max\_\rho -\rho_{center_j}}  
\end{equation}
where $new\_\rho_{centerj}$ denotes the weight value of cluster center $center_{j}$  used for distance optimization, $Max\_\rho$  and $ Min\_\rho$ denote the maximum and minimum density of data points not in \bm{$outc$} respectively.
\begin{equation}\label{eq15}
	S_{new}(x_{i},center_{j})=\sqrt{2(1-exp(-\frac{||x_{i}-center_{j}||^{2}}{\sigma} ))}\times (new\_\rho_{centerj})^{2}
\end{equation}
where $S_{new}(x_{i},center_{j})$ denotes the weighted distance between $x_{i}(x_{i} \notin outc)$  and $center_{j}$. As a result, all the clusters of current iterative and corresponding cluster centers are refreshed. The weighted distance makes data points move closer to cluster centers with relatively smaller density nearby, which is conducive  to the determination of cluster boundaries.
\begin{figure}[H]
	\centering
	\includegraphics[width=8.5cm]{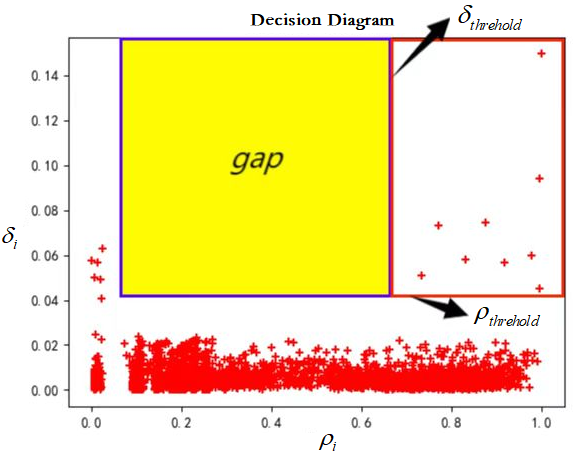}
	\caption{Two-dimensional decision graph based on  $\rho_{i}$ and $\delta_{i}$  .} 
	\label{Fig5}
\end{figure}

For each iteration, the Sum of Squares of Errors (SSE) for all the points are updated. Then, the KMeans iterative process will enter the next refresh cycle of cluster centers until SSE value has little change compared with the previous round. Finally, all the data points except \bm{$outc$} are assigned to the $k$ clusters.

The Parallel Hybrid Clustering Analyzer (\textbf{PHCAL}, \citealt{MA+etal+2022}) combines the advantages of McDPC and KMeans algorithms to ensure the stability and depth of data mining for pulsar candidates. Compared to PHCAL, the clustering process of FMFHC can improve the flexibility of determining initial cluster centers on data sets with irregular shape distribution and get more stable clustering and outliers.

\subsection{Data Partition Strategy}

The statistics show that the FAST 19-beam receiver can provide more than a million candidates per night, mentioned in (\citealt{liu+etal+2021}) and (\citealt{yin+etal+2022}). To improve the time performance of FMFHC, it is essential to the study of the parallel implementation of FMFHC based on the models e.g. MPI (Message Passing Interface) and SparkCore, etc. For this reason, the sliding window based data partition strategy for candidate data streams (\citealt{MA+etal+2022}) is adopted on the basis of data structure. As can be seen in Fig.6, the window size of each round is fixed to Batchsize = $w$. Then, a relatively complete set is formed by selecting appropriate samples from the actual pulsars of various type, which will be added to the block to be detected (shadow areas) at a specific ratio $(v : w)$ in each round. According to the clustering results in every block, the clusters whose pulsar sample proportion is greater than a certain threshold (e.g. 50 percent) will be regarded as pulsar data areas and entered into a unified list for further validation. In addition, the outliers screened out before clustering should be further determined whether they are special pulsars. The sliding window mode enables each data sample to appear in two or more blocks with multiple data distributions, so it becomes possible for some data points classified incorrectly in some blocks being identified correctly in other blocks. Note that, the specific parallelization schemes are not discussed in this work.
\begin{figure}[H]
	\centering
	\includegraphics[width=12cm]{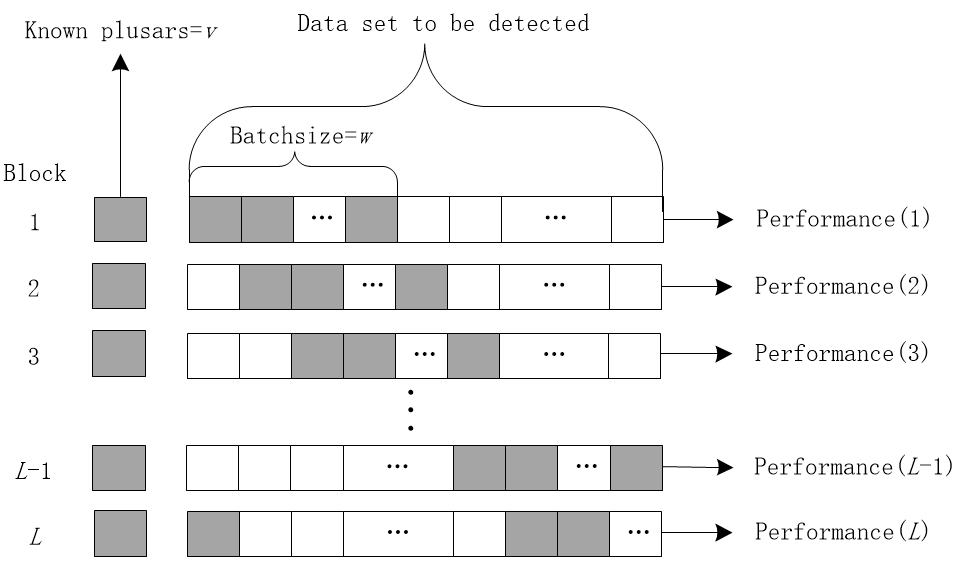}
	\caption{ Data partition based on sliding window  .} 
	\label{Fig6}
\end{figure}

\subsection{Time Complexity Analysis}

As a combination of clustering idea of density-based and partition-based methods, the serial clustering process of FMFHC is more complex than KMeans and DPC algorithms. However, this deficiency can be made up as much as possible by using the reasonable parallelization method. Table 2 shows the time complexity of serial clustering process of FMFHC and compared with other three common serial algorithms, i.e. KMeans++ (\citealt{arthur+2007}), McDPC (\citealt{wang+2020+mcdpc}) and KNN (\citealt{Peterson+2009}). 
\begin{table}[H]
	\bc
	\begin{minipage}{60mm}
		\caption[]{Time complexity statistics of FMFHC and other serial algorithms}

	\end{minipage}
	\setlength{\tabcolsep}{2.5pt}
	\small
	\begin{tabular}{ccccccccccccc}
		\hline\noalign{\smallskip}
		Algorithm &  \makecell{Parallel mode \\ of FMFHC}& \makecell{Serial mode \\ of FMFHC}& KMeans++ & $McDpc$ &
		KNN & \\
		\hline\noalign{\smallskip}
		\makecell{Time \\ Complexity}& $\lim_{G(p) \to 1} O((G(p)m)^{2})
		$&  $\makecell{O(n^{2}),\\ if\sum_{i=1}^{4}C_{i}<threshold_{1}}$ &  $O(nKTM)$& $O(n^{2})$ &   $O(nM+nD^{a}) $   \\
		\noalign{\smallskip}\hline
	\end{tabular}
	\ec
	\tablecomments{0.86\textwidth}{$T$: the number of iterations; $M$: the characteristic number of elements; $m$: the number of samples of Block(i); $k$: the number of cluster centers; $D$: the nearest neighbor parameter}
\end{table}

Let the total number of samples of data set be n, then: i)The time complexity of KMeans++ is  $O(nkTM)$, which can be simplified to $O(n)$ as $k$, $T$ and $M$ are considered as constants. ii)The time complexity of McDPC based on different density levels is $O(n^{2})$  since the computing complexity of parameters $\rho$     and  $\delta$ is $O(n^{2})$. iii)The complexity of KNN is $O(nM+nD)$   that calculated from the worst case. iv)The serial time complexity of FMFHC without   applying the data partition strategy in Section 3.3 is $O(\sum_{i=1}^{4}nC_{i}L+n^{2}+nKTM)$, where $O(\sum_{i=1}^{4}nC_{i}L)$  denotes the time complexity of original CAE architecture which has four convolution layers,    $O(n^{2})$ denotes the time complexity of multi-density center selection scheme and $O(nkTM)$  denotes the time complexity of the improved cluster center iterative optimization of KMeans. Note that, $C_{i}$  is the time complexity of a single convolution layer $i\ (1 \leq i \leq 4)$  of a sample (that is $C_{i}=O(V_{i}^{2}\times W_{i}^{2}\times Cin_{i}\times Cout_{i} )$, where $V_{i}$  denotes the size of output feature map of convolution layer $i$,   $W_{i}$ denotes the size of convolution kernel of convolution layer $i$, $Cin_{i}$ denotes the number of input channels,  $Cout_{i}$ denotes the number of output channels), and $L$  is the number of training iterations. In addition, the time complexity of the feature fusion process between the 1-D feature arrays (8$\times$ 1) and (64$\times$ 1) is closed to $O(n)$ according to Section 3.1, which could be neglected.

If $\sum_{i=1}^{c}C_{i}$ are small enough (i.e. $\sum_{i=1}^{4}C_{i} \leq threshold_{1}$ ) and $n$  is large enough (i.e. $n > threshold_{2}$ ), the serial time complexity of FMFHC could be simplified to  $O(n^{2})$  , where  $\sum_{i=1}^{4}C_{i}$ , $k$, $T$, $M$ and $L$   are considered as constants. Obviously, this is an ideal state and the complexity value is close to McDPC but higher than KNN and KMeans++. In terms of this premise, the time complexity of parallel mode of FMFHC can be further discussed when using the sliding window based data partition strategy. As a result, the parallel complexity of FMFHC is $O((G(p)m)^{2})$  in the light of Sun-Ni theorem (\citealt{sun+1993}), where $G(p)$  denotes the factor and m denotes the number of samples in a Block(i) and $m \ll n$
. When the number of parallel nodes p tends to a certain threshold (close to the total number of divided blocks) and the communication delay tends to be ignored, the complexity value is simplified to $O(m^{2})\ (G(p)\rightarrow 1)$. Therefore, if the communication overhead is very low even negligible, the time complexity of the parallel mode of FMFHC is significantly lower than that of its serial version in theory, which will be verified in practice later. In addition, the speedup $(S_{p})$ and parallel efficiency $(E_{p})$ for the parallel version of FMFHC are defined as follows.
\begin{equation}\label{eq16}
	S_{p}=\frac{T_{s}}{T_{p}}=(\frac{O(n^{2})}{\lim_{G(p) \to 1} O((G(p)m)^{2})})^{, m \ll n, \sum_{i=1}^{4}C_i < \text{$threshold$}_1}
\end{equation}
\begin{equation}\label{eq17}
	E_{p}=\frac{S_{p}}{p}
\end{equation}
where $T_{s}$ is the serial running time of FMFHC, $T_{p}$ is the running time of the parallel mode of FMFHC under $p$ parallel nodes. In theory, the performance of parallel mode of FMFHC will remain consistent for different data sizes and hardware resources, depending following two conditions: i) The $p$ value is close to a sufficiently big threshold; ii) The ratio of communication delay in total running time is small enough to be neglected.

\section{Experiments and Results}
\subsection{Datasets and Evaluation Metrics}

Our algorithm was tested on both HTRU (High Time Resolution Universe Survey) 2 and FAST data sets. HTRU2 is an open telescope data set describes a sample of pulsar candidates collected during the HTRU Survey, which consists of 16259 non-pulsar samples and 1639 pulsar samples. It is widely adopted to evaluate the performance of ML based classification algorithms. (\citealt{Lyon+etal+2016}) made the HTRU2 data set available, which has been uploaded on website\footnote[1]{https:// figshare.com/articles/dataset/HTRU2/3080389/1}. The class imbalance ratio of HTRU2 is 9.92 : 1. It is widely adopted to evaluate the performance of ML based classification algorithms. Another FAST data set is obtained from the actual observation data of FAST (CRAFTS). The CRAFTS database is uploaded on website\footnote[2]{http://groups.bao.ac.cn/ism/CRAFTS/202203/t$20220310 \_683697$.html} . In FAST data set, 157616 candidates with pfd files were collected from the survey, among which 78 were pulsar samples and 157538 RFI samples. The class imbalance ratio of FAST data set is 2019.71:1. Table 3 shows the basic information of both experimental data sets.
\begin{table}[H]
	\bc
	\begin{minipage}[]{100mm}
		\caption[]{  Basic information of both data sets used in this study
			\label{tab2}}\end{minipage}
	\setlength{\tabcolsep}{2.5pt}
	\small
	\begin{tabular}{ccccccccccccc}
		\hline\noalign{\smallskip}
		Data set &  Samples & Pulsars& Non-pulsars & $R^{a}$  \\
		\hline\noalign{\smallskip}
		HTRU2& 17898&  1639 &  16259 &   9.92:1   \\
		FAST data & 157616 &78&157538&2019.71:1 \\
		\noalign{\smallskip}\hline
	\end{tabular}
	\ec
\tablecomments{0.86\textwidth}{$R$: the class imbalance ratio of non-pulsars to pulsars}
\end{table}

The evaluation metrics adopted for performing candidate classification usually are Precision, Recall, and F1-Score. Table 4 shows the confusion matrix of the classification. Precision means the proportion of actual pulsar samples properly classified in all the candidates which are classified as positive, and Recall is the proportion of actual pulsars correctly classified.  Precision and Recall are often inversely proportional (When Precision is high, Recall is usually low), so F1-Score can be used to reconcile this pair of  metrics. Combined with the data partition strategy in Section 3.3, the overall performance metrics i.e. $Precision_{overall}$, $Recall_{overall}$, and $F1-Score_{overall}$ are defined as follows:
\begin{equation}\label{eq18}
	Precision_{overall}=\frac{1}{L}(\sum_{l=1}^{L}\frac{TP_{l}}{(TP_{l}+FP_{l}}))
\end{equation}
\begin{equation}\label{eq19}
	Recall_{l}=\frac{TP_{l}}{TP_{l}+FN_{l}} (1\leq l\leq L)
\end{equation}
\begin{equation}\label{eq20}
	F1-Score_{overall}=\frac{1}{L}(\sum_{l=1}^{L} \frac{2\times Precision_{l}\times Recall_{l}}{Precision_{l}+Recall_{l}})
\end{equation}
where $TP_{l}$, $FP_{l}$ and $FN_{l}$ respectively denote the number of True Positive, False Positive and False Negative in $Block(l)$, and $L$ is the total number of divided data blocks.
\begin{equation}\label{eq21}
	Recall_{overall}=\frac{UTP}{TP+FN}
\end{equation}
where $UTP=TP_{1}\cup TP_{2}\cup TP_{3}\cdot \cdot \cdot TP_{L}
$   means the union of identified pulsar samples in each data block, $TP$ and $FN$  respectively denote the total number of True Positive and False Negative in entire data set. Note that, once a pulsar sample has been correctly identified in a $Block(l)$, it will be counted to $TP$.
\begin{table}[H]
	\bc
	\begin{minipage}[]{100mm}
		\caption[]{ Confusion matrix\label{tab3}}\end{minipage}
	\setlength{\tabcolsep}{30pt}
	\small
	\begin{tabular}{ccccccccccccc}
		\hline\noalign{\smallskip}
		Actual category\\Predicted Results &  Positive & Negative  \\
		\hline\noalign{\smallskip}
		True& True positive(TP)&  False Negative(FN)   \\
		False data & False positive(FP) &True Negative(TN) \\
		\noalign{\smallskip}\hline
	\end{tabular}
	\ec
\end{table}

\subsection{Clustering effect test using the HTRU2 data set}
\begin{figure}[H]
	\centering
	\includegraphics[width=10cm]{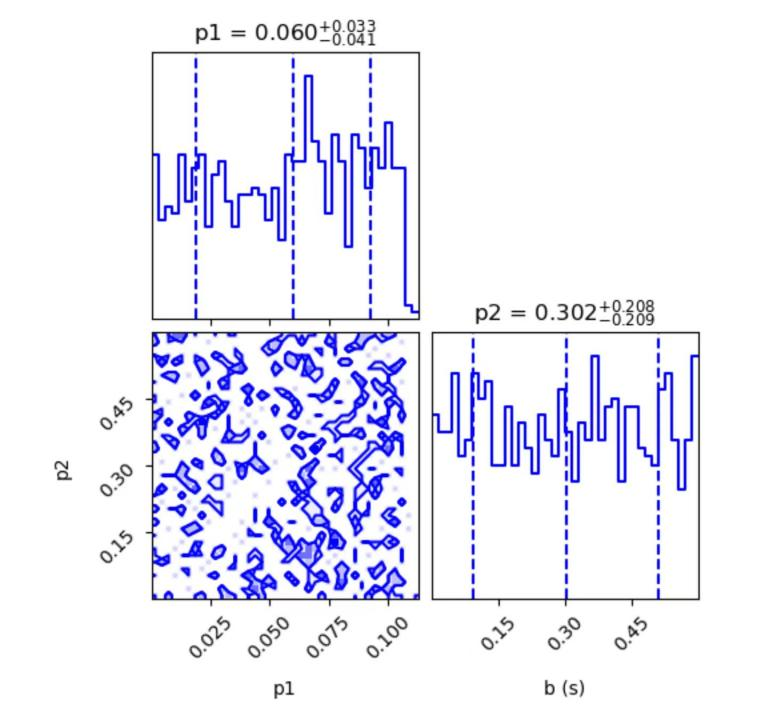}
	\caption{The $\rho_{threhold}$  and $\delta_{threhold}$  triangle plot. Note that,$p1$ denotes  $\rho_{threhold}$ 	and $p2$ denotes  $\delta_{threhold}$  .} 
	\label{Fig7}
\end{figure}

To validate the clustering result of our model, it has been experimented on HTRU2 and compared with other single-modal candidate signal classifiers, including supervised and unsupervised learning algorithms implemented on HTRU2 in the mentioned literature. The data pre-processing was carried on firstly. According to the multi-modal fusion strategy in Section 3.1, all the new candidates on HTRU2 were formed by the fusion of the original 1-D feature arrays and related 2-D TVP arrays. Note that, the 2-D TVP arrays were extracted from other selected pfd files with very similar characteristics to corresponding HTRU2 samples. Moreover, 800 pulsar samples and 4259 non-pulsar samples were used for DCA algorithm training. Next, 1600 of the 1639 real pulsar samples of HTRU2 were randomly selected as a pulsar set $s$, while the remaining 39 pulsar samples were randomly dispersed to the non-pulsar samples of HTRU2 to form the data set to be detected. In terms of the data partition strategy in Section 3.3, the sliding window size was set to Batchsize = 2 where the unit-size was 1161, then the entire data set to be detected was divided into $(t_{1},t_{2},\cdot \cdot \cdot,t_{13},t_{14})$ based on unit-size, where $t_{1},t_{2},...,t_{13}=1161$  but $t_{14}=1205$. Consequently, the experimental data set consists of 14 data blocks, including $\left\{ Block(1):[s,t_{1},t_{2}], Block(2):[s,t_{2},t_{3}], \ldots, Block(13):[s,t_{13},t_{14}], Block(14):[s, t_{14}, t_{1}] \right\}$. Each $Block(i)$  was clustered separately, and the clusters with higher pulsar proportions than a certain value(e.g. $\geq$ 50 percent) in the clustering results were selected into the pulsar-like candidate list.

The values of $\rho_{threhold}$  and $\delta_{threhold}$  have a significant influence on the F1-Score, so they are important for clustering effect. During the execution of our algorithm on HTRU2, the values of $\rho_{threhold}$  and $\delta_{threhold}$   were randomly changed over 2000 rounds. Then, the 800 pairs of $\rho_{threhold}$  and $\delta_{threhold}$   (the corresponding values of F1-Score were not less than 0.8) in the 2000 rounds were selected to make actual 2 parameter triangle plot, as shown in Fig.7. The fitting result shows that there is no interaction between them. Moreover, both $\rho_{threhold}$ and $\delta_{threhold}$  can be fine-tuned by using the heuristic methods (\citealt{Wang+etal+2020}).  On HTRU2, by analyzing the K-dist graph we plot, the respective inflection points of $\rho_{i}$ and $\delta_{i} (x_{i}\notin outc)$  and   were easy to be found. Consequently, $\rho_{threhold}$  may be taken from a range of values $ ( \rho_{i}\in [0.2,0.5]) $ to obtain the best results, and it is the same for  $\delta_{threhold}\ ( \delta_{i}\in [0.02,0.075])$, and then the values were be fine-tuned heuristically. For data sets with single local density level, the reasonable thresholds for $\rho_{i}$  and $\delta_{i}$  will result in the inclusion of all obvious cluster centers, based on the DPC’s assumption that data points with higher $\rho$ and higher $\delta$ values should be selected as cluster centers. By such a heuristic method, it is not difficult to find reasonable $\rho_{threhold}$ value $(\rho_{threhold}=0.3)$ and $\delta$ value $(\delta_{threhold}=0.024)$. In addition,  $\rho_{outlier}$ is set to 0.00051 according to Section 3.2.
\begin{table}[H]
	\bc
	\footnotesize
	\begin{minipage}[]{100mm}
		\caption[]{ Classification results with different methods on HTRU2\label{tab4}}\end{minipage}
	\setlength{\tabcolsep}{6pt}
	\small
	\begin{tabular}{ccccccccccccc}
		\hline\noalign{\smallskip}
		Classfication &      Method  & Precision&Recall&F1-score \\
		\hline\noalign{\smallskip}
		&  SVM  &  0.723&0.901&0.789   \\
		Supervised& PNCN&0.923&0.831&0.874 \\
		&Random Forest &0.958&0.891&0.921 \\
		&KNN  &0.952&0.875&0.909 \\
		\noalign{\smallskip}\hline
		&   KMeans++(k=35)  &  0.926&0.747&0.827   \\
		& McDpc&0.592&0.288&0.388 \\
		Unsupervised Semi-supervised & PHCAL &0.946&0.905&0.881 \\
		&the parallel mode Of FMFHC  &\bm{$0.981$}
		 &\bm{$0.988$}&\bm{$0.974$} \\
		\noalign{\smallskip}\hline
	\end{tabular}
	\ec
\end{table}

Table 5 shows the classification performance of FMFHC on HTRU2, compared with other unsupervised/semi-supervised algorithms (including KMeans++ (\citealt{arthur+2007}), McDPC (\citealt{wang+2020+mcdpc}) and PHCAL (\citealt{MA+etal+2022}) and supervised algorithms (including the Random Forest in (\citealt{Chakraborty+2019}), and the KNN, SVM and PNCN in (\citealt{xiao+etal+2020}) implemented on HTRU2. Among all these unsupervised/semi-supervised and supervised algorithms, the parallel mode of FMFHC has the highest Precision (reaches to 98.1 percent), Recall (reaches to 98.8 percent) and F1-Score (reaches to 97.4 percent). Moreover, upon several rounds of the control test that 39 pulsar samples were randomly selected to form the data set to be detected, all the 39 pulsar samples can be detected out (i.e.100 percent) in each round by the parallel mode of FMFHC. 

\subsection{Robustness test using the FAST data}

The FAST data set was used to further verify the robustness and efficiency of FMFHC. Note that, the data pre-processing on FAST data is very similar to that on HTRU2. At first, to change the class imbalance ratio between pulsar and non-pulsar samples in a single block, 1600 known pulsar samples from multiple types were prepared as a known sample set $s^{'}$ and added to this data set. Some of the pulsar samples in $s^{'}$  are known pulsars searched by CRAFTS during synchronization testing, and they can be found in the linked star catalog\footnote[3]{http://groups.bao.ac.cn/ism/CRAFTS/CRAFTS/}. More over, the dispersion measurement (DM) range of these pulsar samples is often set to [2,1000] ($cm^{-3} pc$). The remaining in   $s^{'}$ were collected from foreign surveys such as HTRU. All of these pulsar samples are normal pulsars. According to Section 3.1, all the new candidates in FAST data (containing the above known sample set $s^{'}$ ) were formed by the DCA fusion of the 1-D feature arrays and related 2-D TVP arrays, which were extracted from corresponding pfd files by the feature extraction programs of HTRU2 and PICS. Moreover, the known sample set $s^{'}$  (1600 real pulsar samples) and 10000 non-pulsar samples were used for DCA algorithm training. Next, the sliding window size was also set to Batchsize = 2 and the unit-size was 1251, then the original data set was divided into $(g_{1}, g_{2},\cdot \cdot \cdot,g_{125}, g_{126})$, where  $g_{1}, g_{2},\cdot \cdot \cdot,g_{125}=1251 $ but $ g_{126}=1241$. As a result, the experimental data set consisted of 126 data blocks, i.e.Block(1):[ $s^{'} , g_{1}, g_{2}$], Block(2):[ $s^{'} , g_{2}, g_{3}$],..., Block(125):[ $s^{'} ,g_{125}, g_{126}$], Block(126):[ $s^{'} , g_{126}, g_{1}$]. Note that, the original 78 pulsar samples from FAST data were randomly distributed in $(g_{1}, g_{2},\cdot \cdot \cdot,g_{125}, g_{126})$. The subsequent clustering process is the same as that on HTRU2, refers to Section 3.2. The specific parameters $\rho_{threhold}=0.3$ , $\delta_{threhold}=0.023$ and $\rho_{outlier}=0.00051$ .
\begin{figure}[H]
	\centering
	\includegraphics[width=12cm]{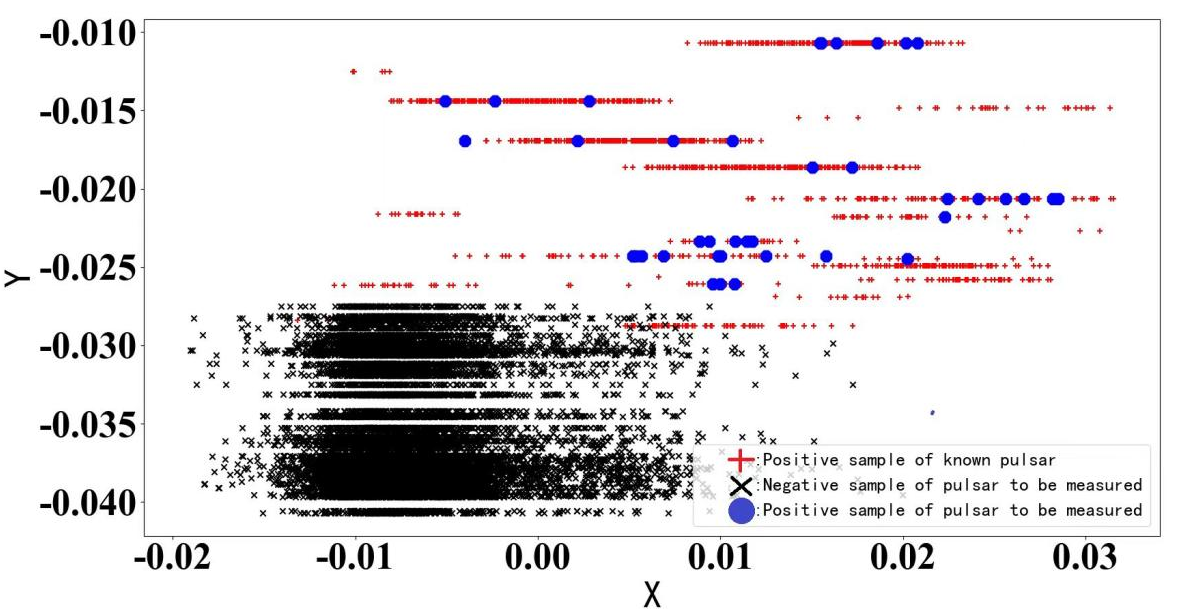}
	\caption{ The clustering effect of FMFHC on FAST data. Note that, the
		input data after multi-modal feature fusion are transformed into the
		arrays with two features X and Y .} 
	\label{Fig8}
\end{figure}

Our algorithm has been experimented by using a Linux cluster environment with seven physical computing nodes with seven Intel 6230 Xeon @ 2.1GHz CPUs with 480 CPU cores(4 Nvidia-GeForce-RTX-2080Ti, 5.3T of total RAM, 3.6P of total disk space). The system is running with Linux 3.10.0-862.el7.x86\_64, Python 3.8, Tensorflow 2 and MPI4py. The experiment results demonstrate that, the highest number of the pulsars identified in a round by the parallel mode of FMFHC achieves 76 of 78, with an average of 75 (Recall of 96.1 percent, Precision of 89.1 percent and F1-score of 92.7 percent), compared with the PICS (Recall of 95 percent) and PICS-ResNet (Recall of 98 percent) in mentioned literature (\citealt{wang2019pulsar}). Fig.8 shows the clustering effect of FMFHC on FAST data.

In addition, the running time data were collected under different number of parallel nodes to further verify the time complexity of FMFHC. Note that, the entire running time of FMFHC refers to the sum of execution time of data import, CAE based feature
extraction, DCA based feature fusion and hybrid clustering, excluding the training time for the CAE and DCA models. Fig.9 shows the average running time of parallel FMFHC with different number of CPU cores. As can be seen from the figure, the average running time decreases as the increment of parallel nodes within limits. When the number of cores reaches 36, it drops to 90.85 seconds.
\begin{figure}[H]
	\centering
	\includegraphics[width=12cm]{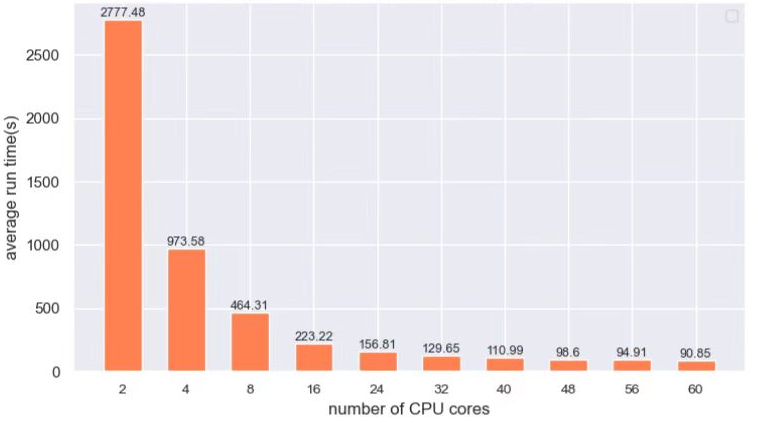}
	\caption{ Average running time of parallel FMFHC based on MPI  .} 
	\label{Fig9}
\end{figure}
\begin{table}[H]
	\bc
	\begin{minipage}[]{100mm}
		\caption[]{Failure modes and codes\label{tab5}}\end{minipage}
	\setlength{\tabcolsep}{2.5pt}
	\small
	\begin{tabular}{ccccccccccccc}
		\hline\noalign{\smallskip}
		Code  &       How fail &  Failure cause& Existence \\
		& & &Yes& No \\
		\hline\noalign{\smallskip}
		301&   Isolated points &  Classification criteria& &$\checkmark$   \\
		302& Collinearity& Signal feature selection&& $\checkmark$\\
		303&Data standardization&Dimension of indicators&&$\checkmark$ \\
		304	&Pattern similarity measure &Data Volume and Variety&$\checkmark$& \\
		305	&   Small pulsar samples clustering  & \makecell{Class imbalance between true \\ pulsars and noise candidates } 
		&	& $\checkmark$ \\
		306	& Fast clustering of large sample data  & \makecell{Cluster center selection and  \\ manual setting of cluster numbers}&$\checkmark$& \\
		\noalign{\smallskip}\hline
	\end{tabular}
	\ec
\end{table}

After analysis, it is concluded that the non-pulsar based transients in FAST data could be roughly classified into other cosmic source signals such as Fast Radio Burst (FRB), RFI, low DM or narrowband signals, and very weak signals, and they can be identified by the signal shapes and signal features. Most of the non-pulsar signals incorrectly identified as pulsars here are broadband RFI. Moreover, the failure modes of clustering methods include following 6 categories, where pattern similarity measure and fast clustering of large sample data still exist for FMFHC, as shown in Table 6.

\section{Discussion}

The overall performance analysis of FMFHC includes two parts: classification performance testing and running time evaluation. On HTRU2, FMFHC can ensure the clustering effect (the highest Precision, Recall and F1-Score) which looks better than other single-modal pulsar candidate classification methods in mentioned literature. On the FAST experimental data collected from CRAFTS, it still performs good (Precision and Recall) compared with the PICS and PICS-ResNet, and its parallel mode significantly reduces the execution time while ensuring the classification performance. It should be explained that HTRU2 is a public dataset specifically designed to test the performance of pulsar candidate classifiers, where each data sample is carefully selected. However, FAST actual observation data are more diverse and complex on data distribution than HTRU2 data. So, the performance of FMFHC on CRAFTS database of FAST does not appear as excellent as that of HTRU2. Consequently, we can believe that FMFHC is effective for high-volume pulsar candidate data streams in actual scenarios. i) a pulsar candidate data stream regardless of its capacity, can be divided into fixed size blocks for parallel processing. ii) it will further promote the discovery of outliers by clustering more meaningful classifications. iii) it still could be improved with the optimization of data partition strategy and relevant parameters. In brief, our algorithm has provided a good theoretical and practical reference for sifting large numbers of pulsar candidate signals obtained from FAST survey.

\section{Conclusion}

A multi-modal hybrid clustering method named FMFHC is presented for large numbers of pulsar candidates in this paper, whose contributions are summarized as follows: 

i)A feature-level fusion scheme based on the DCA algorithm is applied to maximize the separation between pulsar and non-pulsar candidates for large amounts of pulsar candidate data.

ii)A combination of the multi-density peak identification scheme using mixed kernel function for density computing and extension of cluster center iterative optimization scheme of KMmeans is adopted to improve clustering effect for data distribution with multiple shapes and screen out the outliers which could be special pulsars.

iii)The semi-supervised learning mode without large number of training samples and sliding window based data partition strategy are adopted to enhance the efficiency of the overall algorithm and reduce the execution time.

FMFHC is proved to be feasible, but there are still the following shortcomings: i)The enough real data are still needed to further validate our algorithm. ii)Due to limitations in experimental conditions, the actual performance comparison between our algorithm and other advanced parallel algorithms in recent years has not been conducted in an MPI experimental environment. In the future, we plan to connect the proposed algorithm to the pulsar distributed search pipeline based on PRESTO for application test and improving. The program codes of FMFHC are uploaded on website\footnote[4]{https://github.com/You-Ziyi/FMFHC}.

\normalem
\begin{acknowledgements}
	This research is partially supported by the National Key R\&D Program of China (No. 2022YFE0133700), the National Natural Science Foundation of China (No. 12273008, 11963003, 12273007 and 62062025), the National SKA Program of China (No. 2020SKA0110300), the Guizhou Province Science and Technology Support Program (General Project), No. Qianhe Support [2023] General 333, Science and Technology Foundation of Guizhou Province (Key Program, No. [2019]1432), the Guizhou Provincial Science and Technology Projects (No. ZK[2022]143 and ZK[2022]304), the Cultivation project of Guizhou University (No.[2020]76). This work made use of the data from FAST (Five-hundred-meter Aperture Spherical radio Telescope). FAST is a Chinese national mega-science facility, operated by National Astronomical Observatories, Chinese Academy of Sciences. We would like to thank Lyon for providing the publicly available data set and feature extraction scripts, and Zhu W. for providing the feature extraction program of PICS. They are very helpful for our research.
\end{acknowledgements}
  
\bibliographystyle{raa}
\bibliography{mian}

\end{document}